\documentclass[10pt]{article}
\usepackage{amssymb,amsfonts,amsmath,latexsym}
\usepackage{graphics,graphicx,epsf}
\newcommand{\be}{\begin{equation}}
\newcommand{\ee}{\end{equation}}
\newcommand{\bea}{\begin{eqnarray}}
\newcommand{\eea}{\end{eqnarray}}
\newcommand{\ba}{\begin{array}}
\newcommand{\ea}{\end{array}}
\newcommand{\bt}{\begin{tabular}}
\newcommand{\et}{\end{tabular}}

\newcommand{\fr}{\frac}
\newcommand{\ci}{\cite}
\newcommand{\cl}{\centerline}
\newcommand{\bs}{\bigskip}

\newcommand{\vs}{\vspace}

\newcommand{\en}{\eqno}

\newcommand{\bbib}{\begin{thebibliography}}
\newcommand{\ebib}{\end{thebibliography}}

\begin{document}

\begin{flushright}
{\it In memory of my father}\\
\end{flushright}
\bs

\cl{\bf DUALITY AND EFFECTIVE CONDUCTIVITY}
\cl{\bf OF TWO-DIMENSIONAL TWO-PHASE SYSTEMS}

\bs

\cl{\bf S.A.Bulgadaev \footnote{e-mail: sabul@dio.ru}}

\bs
\cl{Landau Institute for Theoretical Physics}
\cl{Chernogolovka, Moscow Region, Russia, 142432}

\bs

\begin{quote}
\footnotesize{
The possible functional forms of the effective
conductivity $\sigma_{eff}$ of the randomly inhomogeneous two-phase system at
arbitrary values of concentrations are discussed.
A new functional equation, generalizing the duality relation,
is deduced for systems with a finite maximal characteristical scale of the
inhomogeneities and its solution is found.
A hierarchical method of the construction  of the model random inhomogeneous
medium is proposed and one such simple model is constructed. Its
effective conductivity at arbitrary phase concentrations is found
in mean field like approximation. The derived formulas
(1) satisfy all necessary inequalities and symmetries, including a dual
symmetry;
(2) reproduce the known formulas for $\sigma_{eff}$ in weakly inhomogeneous
case.
It means that in general $\sigma_{eff}$ of the
two-phase randomly inhomogeneous system may be a nonuniversal function and
can depend
on some details of the structure of the randomly inhomogeneous regions.
The percolation limit of the randomly inhomogeneous two-phase systems
is briefly discussed.}
\end{quote}

\bs

The electrical transport properties of the disordered systems have
an important practical interest. For this reason they are intensively studied
theoretically as well as experimentally. In this region there is one classical
problem about the effective conductivity $\sigma_{eff}$ of inhomogeneous
(randomly or regularly) heterophase system which is a mixture of $N (N \ge 2)$
different phases with different conductivities  $\sigma_i, i =1,2,...,N.$
We confine ourselves here by the simplest case of the two-dimensional heterophase
systems with $N=2.$
Despite of its relative simplicity only some few general exact results have
been
obtained so far. Firstly, there is a general expression for $\sigma_{eff}$
in case of weakly inhomogeneous isotropic medium, when the conductivity
fluctuations $\delta \sigma$ are smaller than an average conductivity
$\langle \sigma \rangle$ ($\delta \sigma \ll \langle \sigma \rangle$)
similar to the analogous expression for the
permitivity of the corresponding dielectric mixture \ci{1}
$$
\sigma_{eff} = \langle \sigma \rangle \left(1 -
\fr{\langle \sigma^2 \rangle - {\langle \sigma \rangle}^2}
{D{\langle \sigma \rangle}^2}\right),
\en(1)
$$
where $D$ is a dimension of the system.
In our simplest case of two-dimensional two-phase system
$\langle \sigma \rangle = x \sigma_1 + (1-x) \sigma_2, \;
\langle \sigma^2 \rangle - \langle \sigma \rangle^2 = x(1-x)(\Delta \sigma)^2,$
where $x$ is a concentration of the first phase,
$\Delta \sigma = \sigma_1 - \sigma_2$,
and the first formula takes a form
$$
\sigma_{eff} =
\sigma_1\left(1 - (1-x)\fr{\Delta \sigma}{\sigma_1} -
\fr{x(1-x) (\Delta \sigma)^2}{2\sigma_1^2}\right) =
$$
$$
\sigma_+\left(1 + (x-1/2)\fr{\Delta \sigma}{\sigma_+} -
\fr{x(1-x) (\Delta \sigma)^2}{2\sigma_+^2}\right),
\en(1')
$$
where $\sigma_+ = (\sigma_1 + \sigma_2)/2.$

Another general formula has been obtained
for the case of a small concentration of one phase (for example, with
a conductivity $\sigma_2$) \ci{1}
$$
\sigma_{eff} = \sigma_1 \left(1 -
(1-x)\fr{2(\sigma_1 - \sigma_2)}
{\sigma_1 + \sigma_2}\right),
\en(2)
$$
where $1-x \ll 1$ is a small concentration of the second phase and a round
form of the inclusions of this phase is suggested.

The further progress in the solution of this problem is connected with
papers \ci{2,3}.  It was shown  there that there is a dual transformation
transforming one phase into another. This transformation allows to find an
exact formula for $\sigma_{eff}$ in case of systems with  equal concentrations
of the phases
$x = x_c =1/2$ \ci{3}
$$
\sigma_{eff} = \sqrt{\sigma_1 \sigma_2}.
\en(3)
$$
This remarkable  formula is very simple and universal since it does not depend
on the type of the inhomogeneous structure of the two-phase system.
For systems with unequal phase concentrations the dual transformation gives
a relation between effective conductivities at adjoint concentrations
$x$ and $1-x$ or in terms of a new variable $\epsilon = x-x_c$
($ -1/2 \le \epsilon \le 1/2$) at $\epsilon$ and $-\epsilon$
$$
\sigma_{eff}(x, \sigma_1, \sigma_2 )
\sigma_{eff} (1-x, \sigma_1, \sigma_2) = \sigma_1 \sigma_2 =
\sigma_{eff}(\epsilon, \sigma_1, \sigma_2 )
\sigma_{eff} (-\epsilon, \sigma_1, \sigma_2).
\en(4)
$$
Eq.(4) means that a product of the effective conductivities at adjoint
concentrations is an invariant. Due to this relation one can consider
$\sigma_{eff}$ only in the regions $x \ge x_c$ ($\epsilon \ge 0$) or
$x \le x_c$ ($\epsilon \le 0$).

In the following papers this dual transformation was generalized on systems
in magnetic field \ci{5,6,7}, polycrystals \ci{3} and heterophase (with $N >2$)
systems \ci{8,9}.
The formula (4) was also checked and improved for some regularly inhomogeneous
systems (of the chess-board type) \ci{10,11}.

However the main interest in this problem has a formula for the effective
conductivity at arbitrary phase concentrations.  Another question appears
naturally in this case:
is this formula also universal or it can depend on the structure of the
two-phase system?  We will give here a general analysis of the possible
functional forms of the effective conductivity in the case of arbitrary
phase concentrations and will show that, due to the dual symmetry,
in some cases the corresponding effective conductivities can be found in
the explicit form. Then we will introduce the finite maximal scale averaging
approximation (FMSA-approximation), deduce in the framework of this
approximation a new equation, connecting the effective conductivities
at different concentrations, and find its solution and its physical meaning.
We will introduce also a new model of the
random two-phase system which gives in the same approximation
an effective conductivity, coinciding with one obtained by series expansion.
All these results
demonstrate that, in general, a formula for the effective conductivity may
be nonuniversal even in the two-phase case. At the end of the paper we
will briefly discuss some peculiarities of the percolation limit.

Let us start our investigation of the isotropic classical random
two-phase system in the case of arbitrary concentrations with a general
analysis.
The basic equation is the Ohm law, connecting the local current
${\bf j}({\bf r})$ and local electric field ${\bf E}({\bf r}),$
$$
{\bf j}({\bf r}) = \sigma ({\bf r}) {\bf E}({\bf r}),
\en(5)
$$
where $\sigma ({\bf r})$ is a local conductivity. It must be supplemented
with the corresponding boundary conditions on the boundaries of two phases
\ci{1}
$$
j^1_n = j^2_n, \quad E^1_t = E^2_t,
\en(6)
$$
where $n,t$ denote the normal and tangent components and $1,2$ correspond
to different phases.
The electric field ${\bf E}({\bf r})$ is a curl-free field
$$
\nabla \times {\bf E}({\bf r}) = 0,
\en(7)
$$
and the current field is a divergenceless field
$$
\nabla \cdot {\bf j}({\bf r}) = 0.
\en(8)
$$
The effective conductivity $\sigma_{eff}$ of the isotropic system can be
defined as
a proportionality coefficient between averaged values of ${\bf j}$ and ${\bf E}$
over the area of the system
$$
\bar {\bf j} = \sigma_{eff} \bar {\bf E},
$$
$$
\bar {\bf j} \equiv \int  {\bf j}({\bf r}) d{\bf r}/S, \quad
\bar {\bf E} \equiv \int  {\bf E}({\bf r}) d{\bf r}/S
\en(9)
$$
where $S$ is an area of the system.
Due to the linearity of the defining equations an effective conductivity
of the random systems must be a homogeneous function of degree one
of $\sigma_i, i=1,...,N.$
In our case $N=2$ and it is convenient to use instead of $\sigma_i, i=1,2$
another combinations of $\sigma_i$: $\sigma_{\pm} = (\sigma_1 \pm \sigma_2)/2.$
Then, introducing a new variable $z=\sigma_{-}/ \sigma_+,\; (-1 \le z \le 1),$
an effective conductivity can be represented in the following, symmetrical
relatively to both phases, form
$$
\sigma_{eff}(\epsilon, \sigma_+, \sigma_- ) =
\sigma_{+} f(\epsilon, \sigma_-/ \sigma_+) =
\sigma_{+} f(\epsilon, z),
\en(10)
$$
where $\sigma_{eff}(\epsilon, \sigma_+, \sigma_- )$ and $f(\epsilon, z)$
must have the next boundary values
$$
\sigma_{eff}(1/2, \sigma_+, \sigma_- ) = \sigma_1, \quad
\sigma_{eff}(-1/2, \sigma_+, \sigma_- ) = \sigma_2,
$$
$$
f(1/2, z) = 1+z, \quad f(-1/2, z) = 1-z, \quad f(\epsilon, 0) =1.
\en(10')
$$
The duality relation, being a consequence of a duality of gradient and tangent
fields in two dimensions \ci{3}, takes in these variables the form
$$
f(\epsilon, z) f(-\epsilon, z) = 1-z^2,
\en(11)
$$
from which it follows that at critical concentration $\epsilon = 0$
$$
f(0, z) = \sqrt{1-z^2}.
\en(3')
$$
Strictly speaking, this form of a duality relation is also a consequence
of another exact relation for the effective conductivity,
taking place at arbitrary concentrations for systems with
the similar random structures of both phases of the system,
$$
\sigma_{eff}(\epsilon, \sigma_1, \sigma_2 ) =
\sigma_{eff} (-\epsilon, \sigma_2, \sigma_1).
\en(12)
$$
It means that the effective conductivity of the random two-phase system
must be invariant under substitution of these phases
($\sigma_1 \longleftrightarrow \sigma_2$) with the corresponding
change of their concentrations $x \longleftrightarrow 1-x$ (or $\epsilon
\to -\epsilon$). In the new variables it means that
$$
f(\epsilon, z) = f(-\epsilon, -z),\quad f(-\epsilon, z) = f(\epsilon, -z).
\en(13)
$$
For this reason a duality relation can be written also in the form
$$
f(\epsilon, z) f(\epsilon, - z) = 1-z^2.
$$
It follows from (13) that the even ($f_s$) and odd ($f_a$) parts of
$f(\epsilon,z)$ relatively
to $\epsilon$  coincide with the even ($f^s$) and odd ($f^a$) parts of
$f(\epsilon,z)$
relatively to $z$
$$
f^a(\epsilon,z) \equiv \fr{1}{2} \left(f(\epsilon,z) - f(\epsilon,-z)\right)
= f_a(\epsilon,z)
\equiv \fr{1}{2} \left(f(\epsilon,z) - f(-\epsilon,z)\right),
$$
$$
f^s(\epsilon,z) \equiv \fr{1}{2} \left(f(\epsilon,z) + f(\epsilon,-z)\right)
= f_s(\epsilon,z)
\equiv \fr{1}{2} \left(f(\epsilon,z) + f(-\epsilon,z)\right).
\en(14)
$$
The simplest way to satisfy (13) corresponds to the
following functional form of $f(\epsilon,z)$
$$
f(\epsilon,z) = f(\epsilon z, \epsilon^2, z^2)
\en(15)
$$
One can conclude from (15) that in this case an expansion of $f(\epsilon,z)$
near the
point $\epsilon = z = 0$ does not contain terms linear in $\epsilon$ and $z$
separately.
Moreover, one can reduce to the form (15) any function satisfying (13), since
every combination of functions of two variables, odd relatively to the
inversion of their
arguments, can be made even by multiplying or dividing them by $\epsilon z$
$$
F(-\epsilon, z) = - F(\epsilon, z) = F(\epsilon, -z),
\quad F(\epsilon, z) =
\epsilon z \Phi(\epsilon^2, z^2),
$$
$$
\Phi(\epsilon^2, z^2) = (\epsilon z)^{-1} F(\epsilon, z).
$$
Analogously, the odd part $f_a$ can be represented in the form
$$
f_a(\epsilon,z) = 2\epsilon z \Phi(\epsilon, z)
\en(16)
$$
where $\Phi$ is an even function of $\epsilon$ and $z$ (the coefficient $2$
in front of $\epsilon z$ is chosen for further convenience).

The duality relation (11) in general case is not enough for the complete
determination of $f$, it only connects $f_a$ and $f_s$
$$
f_s^2 - f_a^2 = 1-z^2.
\en(17)
$$
It means that $f_a$ and $f_s$ considered at fixed $z$ as the functions
of $\epsilon$ satisfy to hyperbolic relation with a constant depending on
$z.$ The relation (17) allows to express $f(\epsilon,z)$ through its
even or odd parts
$$
f(\epsilon,z) =
f_a + \sqrt{f_a^2 + 1-z^2} =
f_s \pm \sqrt{f_s^2 - 1 + z^2}.
\en(18)
$$
For this reason it is enough to know only one of these two parts.
Usually one prefers to choose an antisymmetric part as more simple.
It follows from (1')
that in the weakly inhomogeneous case the odd part has the simplest form
(16) with $\Phi=1$
$$
f_a(\epsilon,z) = 2 \epsilon z.
\en(19)
$$
As is well known, the substitution of (19) into (18) gives an effective conductivity in the
effective medium (EM) approximation \ci{4}
$$
\sigma_{eff}(\epsilon,z) = \sigma_+ \left[2 \epsilon z +
\sqrt{(2 \epsilon z)^2 + 1 - z^2}\right],
\en(20)
$$
This expression, being continued on arbitrary concentrations
$\epsilon = x-1/2$ and inhomogeneities $z,$ reproduces in the corresponding
limits both formulas (1') and (2).

Though at first sight a duality relation is not enough for finding $f$,
usually systems with such symmetry have some additional hidden properties,
permitting to obtain more information about function under question.
Moreover, in some cases they can help to solve problem exactly
(see, for example, \ci{12}). Having this in mind, we will try to investigate
the duality relation in more details.
Since a homogeneous limit $z \to 0$ is a regular point of $f$ it will be very
useful to consider a series expansion of $f$ in $z$
$$
f(\epsilon, z) = \sum_0^{\infty} f_k(\epsilon) z^k/k!,
\en(21)
$$
where due to boundary conditions (10')
$$
f_0 = 1, \quad f_1(\epsilon) = 2 \epsilon.
\en(22)
$$
For every fixed
$z \; (0 \le z \le 1)$ a function $f$ must be a monotonous function
of $\epsilon.$
Substituting the expansion (21) into (11) one obtains
the following results:

(1) in the second order in $z$ it reproduces the universal formula (1'),
thus the latter can be considered as a consequence of the duality relation;

(2) in general case there are the recurrent relations between $f_{2k}$ and
$f_{2k-1},$ corresponding to the connection (17);

(3) $f_{2k+1}(\epsilon)$ are odd polynomials of $\epsilon$ of degree
$2k+1$ and
$f_{2k}(\epsilon)$ are even polynomials of $\epsilon$ of degree $2k$
in agreement with (15).

Taking into account boundary conditions (10') and an exact value
(3'), one can show that the coefficients $f_{k}$ must have the next form
$$
f_{2k+1}(\epsilon) = \epsilon (1-4\epsilon^2) g_{2k-2}(\epsilon),
\quad k \ge 1,
$$
$$
f_{2k}(\epsilon) = (1-4\epsilon^2) h_{2k-2}(\epsilon),
\quad k \ge 1,
\en(23)
$$
where $g_{2k-2}$ and $h_{2k-2}$ are some even polynomials of the corresponding
degree and free terms of $h_{2k-2}$ coincide with the coefficients in the
expansion of (3')
$$
\sqrt{1-z^2} = 1-z^2/2 -z^4/8 -z^6/16 -z^8/128 -z^{10}/256 + ...
\en(24)
$$
It follows from (23) that $f_3$ is completely determined up to overall
factor number $g_0.$ Since $f_4$ is determined through lower
$f_k \quad (k=1,2,3)$
$$
f_4 = 4f_1 f_3 - 3f_2^2 = (1-4\epsilon^2)[(8g_0 + 12)\epsilon^2 - 3],
\en(25)
$$
it is also determined by the coefficient $g_0.$ The case
$g_0 = 0 = g_{2k+1} \; (k>1)$ corresponds
to the EM approximation and in this case
$f_{2k}(\epsilon) \sim (1-4\epsilon^2)^k$.
Thus we see that in general case the arbitrariness of $f$ is strongly reduced
by boundary conditions and known exact value and that the third and fourth
orders are determined only up to one constant. Then one can suppose that
any additional information about function $f$ can determine this constant
or even the whole function.
For this reason one can ask: what kind of functions can
satisfy the duality relation (11)?
In order to answer on this question
it is convenient in the case $z \ne 1$ to pass from $f$ to
$\tilde f = f/\sqrt{1-z^2}.$ Then
$$
\tilde f(\epsilon, z) \tilde f(-\epsilon, z) = 1 =
\tilde f(\epsilon, z) \tilde f(\epsilon, -z).
\en(11')
$$
The duality relation gives some constraints
on the possible functional form of $\tilde f(\epsilon,z).$
For example, assuming a functional form (15), one can write out
a simple function:
$$
\tilde f(\epsilon,z) = \exp(\epsilon z \phi(\epsilon,z)),
\en(26)
$$
where $\phi(\epsilon, z)$ is some even function of its arguments.
It is easy to see that it automatically satisfies eq.(11').
Let us now consider two simple cases, when (a) $\phi(\epsilon, z)$ depends
only on $z,$ (b) it depends only on combination $\epsilon z.$
Expanding the corresponding functions $\tilde f$ in series one can check
after some algebra
that  one can now determine all polynomial coefficients unambiguously!
Another way to see this is to apply boundary conditions directly to the
function (26). In the case (a) one obtains
$$
\phi(z) = 1/z \ln \fr{1+z}{1-z}, \quad
\tilde f (\epsilon,z) =  \left(\fr{1+z}{1-z}\right)^{\epsilon}.
\en(27)
$$
In case (b), when $\phi$ depends only on combination $\epsilon z$,
one finds
$$
\phi(\epsilon z) = \fr{1}{2\epsilon z} \ln \fr{1 + 2\epsilon z}{1-2\epsilon z},
\quad \tilde f(\epsilon,z) =
\left(\fr{1+ 2\epsilon z}{1-2\epsilon z}\right)^{1/2}.
\en(28)
$$
Series expansions of (27) and (28) coincide exactly with the corresponding
expansions mentioned above.
Unfortunately, analyzing an expansion of (26) under an assumption of regular
behaviour of the coefficients $\phi_k(\epsilon)$ from a series expansion of
$\phi(\epsilon,z)$
$$
\phi(\epsilon,z) = \sum_0^{\infty} \phi_k(\epsilon) z^{2k}/k!
$$
in $\epsilon$
$$
\phi_k(\epsilon) = \sum_0^{\infty} \phi_{kl}(\epsilon)^{2l}/l!,
$$
one can show that now again the
boundary conditions (10') (or (23)) cannot define all coefficients completely.
For example, $f_3$ and $f_4$ contain one free parameter $\phi_{10}:$
$g_0 = 6(\phi_{10} -1).$ It is very important to find any other solutions.
Below we will present two simple models having the effective conductivity
just of two  forms found above.

Firstly, we will obtain an additional equation for effective conductivity
using a new composite method, which we will call as a finite maximal
scale averaging approximation. In framework of this method one divides an averaging
procedure on two steps. This approximation can be considered as some modification
of the mean field approximation and it is applicable to the inhomogeneous
systems with a finite maximal scale of the inhomogeneities.
Below in this section we will use temporally, for clarity,
the concentration variable $x$ instead of $\epsilon.$
Let us consider two-dimensional
two-phase randomly inhomogeneous system.
It is easy to see that, in general, its effective conductivity will depend
on a scale $l$ of a region over which an averaging is done. This takes place
due to the possible existence of different characteristic scales in the
inhomogeneous medium. In the most general case there will be a whole spectrum
of these characteristic scales. This spectrum can be very different from
discrete finite till continuous infinite and will define the inhomogeneity
structure of the system. For this reason this spectrum can depend on
concentration $x.$ It is obvious that the effective conductivity
of the system can depend on this spectrum as a whole as well as on
which area an averaging is fulfilled over.
Suppose, for the simplicity, that the randomly inhomogeneous structure of
our system has the scale spectrum with a maximal scale $l_m(x),$ which is
finite for all $x$ in the region $1 \ge x > 1/2$ (or $1-x$ in the region
$0 \le 1-x < 1/2$). Let us assume that we know an
exact formula for $\sigma_{eff}(x,z)$ of this system, which is applicable from
scales $l > l_m.$
It means that this formula for $\sigma_{eff}(x,z)$ takes place after the
averaging over regions with a mean size $l \gtrsim l_{m}$ and does not
change for all larger scales $l \gg l_m.$
Now consider a square lattice
with the squares of length $l_L \gg l_{m}$ and suppose that they have the
effective conductivities corresponding to different values of the
concentrations $x_1$ and $x_2$ with equal probabilities $p = 1/2$ (see Fig.1)

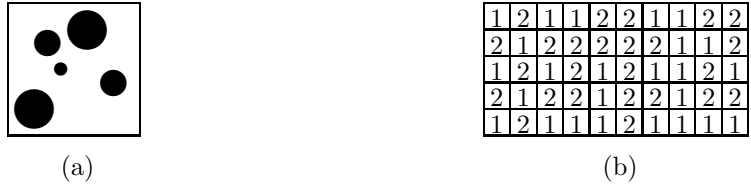
\begin{figure}
\begin{picture}(250,120)
\put(45,20){\line(0,1){50}}
\put(45,20){\line(1,0){50}}
\put(95,20){\line(0,1){50}}
\put(45,70){\line(1,0){50}}
\put(55,30){\circle*{15}}
\put(60,55){\circle*{10}}
\put(75,60){\circle*{20}}
\put(85,40){\circle*{10}}
\put(65,45){\circle*{5}}
\put(65,5){(a)}
\put(150,0){%
\begin{picture}(100,50)%
\multiput(75,20)(10,0){11}%
{\line(0,1){50}}
\multiput(75,20)(0,10){6}%
{\line(1,0){100}}
\multiput(77.5,21)(0,20){3}{1}
\multiput(77.5,31)(0,20){2}{2}
\multiput(87.5,31)(0,20){2}{1}
\multiput(87.5,21)(0,20){3}{2}
\multiput(97.5,21)(0,20){3}{1}
\multiput(97.5,31)(0,20){2}{2}
\multiput(107.5,31)(0,10){3}{2}
\multiput(107.5,21)(0,40){2}{1}
\multiput(117.5,51)(0,10){2}{2}
\multiput(117.5,21)(0,10){3}{1}
\multiput(127.5,21)(0,10){5}{2}
\multiput(137.5,21)(0,20){3}{1}
\multiput(137.5,31)(0,20){2}{2}
\multiput(147.5,21)(0,10){5}{1}
\multiput(157.5,31)(0,10){2}{2}
\multiput(157.5,21)(0,30){2}{1}
\multiput(167.5,31)(0,20){2}{2}
\multiput(167.5,21)(0,20){2}{1}
\put(167.5,61){2}
\put(157.5,61){2}
\put(120,5){(b)}
\end{picture}}
\end{picture}
\caption{\small (a) An elementary square of the model
with round inclusions, (b) a lattice of the model, the numbers 1,2 denotes
squares with the corresponding concentrations.}
\end{figure}

After the averaging over the scales $l \gtrsim l_L$ one must obtain on much
larger scales $l \gg l_{L}$ the same effective conductivity,
but corresponding to another concentration $x=(x_1 + x_2)/2.$
This is possible due to the similar random structure of different squares
and due to the conjectures that in this
model: (1) there are only two maximal characteristic scales
$l_m(x_i) \equiv l_i \quad (i=1,2)$ and the lattice square size
$l_L \gg l_i, l_m(x)$, (2) the averaging procedures over these scales do not
correlate (or weakly correlate) between themselves. Thus for compatibility
all concentrations must be out of small region around critical concentration
$x_c,$ where $l_i$ or $l_m(x)$ can be very large. We will call further
this set of the conjectures the finite maximal scale averaging approximation
(FMSA approximation). It can be implemented for systems with compact
inhomogeneous inclusions with finite $l_m$.
From the other side the effective conductivity
on scales $l \gg l_L$ must be determined  by the universal
Keller -- Dykhne formula (3). Thus we obtain the next functional equation
for the effective conductivity, connecting $\sigma_{eff}(x,z)$ at different
concentrations,
$$
\sigma_{eff}(x,z) = \sqrt{\sigma_{eff}(x_1,z) \sigma_{eff}(x_2,z)},\quad
x=(x_1+x_2)/2, \quad (x_i,x \ne x_c).
\en(29)
$$
It must be supplemented by the boundary conditions (10').
This equation can be considered as a generalization of the duality relation (4),
the latter being a particular case of (29) at $x_1 + x_2 = 1.$
It follows from (29) that, due to the exactness of the duality relation,
it really works for all concentrations $x,$ except maybe of small region
near $x=x_c$ and $z=1$ (this region corresponds to the singular region of
the percolation problem, see also below a discussion of the percolation limit).
One can easily check that $\sigma_{eff}$ in low concentration limit (2)
satisfies (29). Moreover, one can find an exact solution of this
equation. It has an exponential form
with a linear function of $x$ (or $\epsilon$)
$$
\sigma_{eff}(x,z) = \sigma_{1} \exp(ax + b),
\en(30)
$$
where the constants $a,b$ can be determined from the boundary conditions
$$
a = - b, \quad \exp b = \sigma_2/\sigma_1.
\en(30')
$$
Substituting these coefficients into (30) one obtains
$$
\sigma_{eff}(x,z) = \sigma_{1} \left(\sigma_2/\sigma_1\right)^{(1-x)}
= \sigma_+\sqrt{1-z^2}\left(\frac{1+z}{1-z}\right)^{\epsilon}.
\en(31)
$$
The solution (31) satisfies all symmetry relations (4,11',13) and can be
represented in the exponential form
$$
\sigma_{eff}(x,z) =
\sigma_+\sqrt{1-z^2}\exp\left(\epsilon \ln \fr{1+z}{1-z}\right),
\en(31')
$$
which exactly coincides with the case (a) from (27) with
$\phi(\epsilon,z) = \fr{1}{z} \ln \fr{1+z}{1-z}$.
Its odd part has a form
$$
\tilde f_a(\epsilon,z) = \sinh (\epsilon z \phi(\epsilon,z)),
\en(32)
$$
and satisfies (17).

It is interesting to note that the form of the solution (31) means that
in the considered approximation one has effectively an averaging  of the
$\log \sigma$ since it can be represented as
$$
\log \sigma_{eff} = \langle \log \sigma \rangle =
x \log \sigma_1 + (1-x)\log \sigma_2.
\en(33)
$$
This coincides with a note made in \ci{3} for the case of equal concentrations
$x=1/2$ and with analogous results obtained later for random system in
the theory of the localization \ci{13}.
One can check that (31) reproduces
in the weakly inhomogeneous limit the universal Landau -- Lifshitz expression
(1). In the low concentration limit of the second phase it gives
$$
\sigma_{eff}(x,z) = \sigma_1(1 + (1-x)\log \fr{1-z}{1+z} + ...),
\quad 1-x \ll 1,
\en(34)
$$
what coincides with (2) in the weakly inhomogeneous case.
Note that the expansion (34) contains the coefficients diverging in the limit
$|z| \to 1.$  Such behaviour of the coefficients denote the existence of a
singularity in this limit (see below a discussion of this percolation limit).

Now we will show that an existence of the different functional forms for the
effective conductivity denotes a nonuniversal character of this value for
two-phase systems with different random structures.
We will construct  a model of two-dimensional isotropic randomly
inhomogeneous two-phase system, using the composite method introduced above,
and find a mean field like expression for its effective
conductivity $\sigma_{eff}(x,z)$ in case of arbitrary phase concentrations $x.$
The derived formula satisfies again to all necessary symmetries, including
a dual one, and coincides with the example case (b) from (28), realizing
the second variant, when $\tilde f(x,z)$ depends
(in this approximation) only on one combination of variables $\epsilon z$
and this dependence
is described by the  function analytical at small values of this variable.

Let us consider the following two-dimensional model. There is a simple
square lattice with the squares consisting of a random layered mixture of two
conducting phase with constant conductivities $\sigma_i, i = 1,2$ and
the corresponding concentrations $x$ and $1-x.$  A schematic picture of such
square is given in Fig.2.

\begin{figure}
\begin{picture}(250,120)
\put(50,20){\line(1,0){50}}
\put(50,20){\line(0,1){50}}
\put(100,20){\line(0,1){50}}
\put(50,70){\line(1,0){50}}
\put(60,20){\line(0,1){50}}
\put(63,20){\line(0,1){50}}
\put(80,20){\line(0,1){50}}
\put(90,20){\line(0,1){50}}
\multiput(66.5,45)(5,0){3}{\circle*{1}}
\multiput(61.5,23)(0,3){16}{\circle*{2}}
\multiput(83.3,23)(0,3){16}{\circle*{2}}
\multiput(86.6,23)(0,3){16}{\circle*{2}}
\put(65,5){(a)}
\put(150,0){%
\begin{picture}(100,50)%
\multiput(75,20)(10,0){11}%
{\line(0,1){50}}
\multiput(75,20)(0,10){6}%
{\line(1,0){100}}
\put(120,5){(b)}
\multiput(78,25)(0,20){3}{\line(1,0){3}}
\multiput(80,33)(0,20){2}{\line(0,1){3}}
\multiput(88,35)(0,20){2}{\line(1,0){3}}
\multiput(90,23)(0,20){3}{\line(0,1){3}}
\multiput(98,25)(0,20){3}{\line(1,0){3}}
\multiput(100,33)(0,20){2}{\line(0,1){3}}
\multiput(110,33)(0,10){3}{\line(0,1){3}}
\multiput(108,25)(0,40){2}{\line(1,0){3}}
\multiput(120,53)(0,10){2}{\line(0,1){3}}
\multiput(118,25)(0,10){3}{\line(1,0){3}}
\multiput(130,23)(0,10){5}{\line(0,1){3}}
\multiput(138,25)(0,20){3}{\line(1,0){3}}
\multiput(140,33)(0,20){2}{\line(0,1){3}}
\multiput(148,25)(0,10){5}{\line(1,0){3}}
\multiput(160,33)(0,10){2}{\line(0,1){3}}
\multiput(158,25)(0,30){2}{\line(1,0){3}}
\multiput(170,33)(0,20){2}{\line(0,1){3}}
\multiput(168,25)(0,20){2}{\line(1,0){3}}
\put(170,63){\line(0,1){3}}
\put(160,63){\line(0,1){3}}
\end{picture}}
\end{picture}
\caption{\small (a) An elementary square of the model with a
vertical orientation, the dotted regions denote layers of the second phase;
(b)  a lattice of the model, the small lines on the
squares denote their orientations.}
\end{figure}
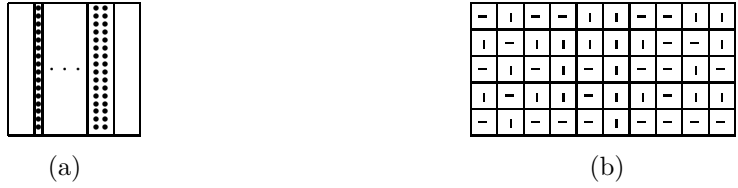

The layered structure
of the squares means that the squares have some preferred direction,
for example
along the layers. Let us suppose that the directions of different squares
are randomly oriented (parallelly or perpendicularly) relatively to the
external electric field, which is directed along $x$ axis.
In order for system to be isotropic the probabilities of the parallel and
perpendicular orientations of squares must be equal or (what is the same)
the concentrations of the squares with different orientations must be equal
$p_{||} = p_{\perp} = 1/2.$

Such lattice can model a random system consisting from mixed phase regions,
which can be roughly represented on the small macroscopic scales as randomly
distributed plots with the effective "parallel" and "serial" connections
of the layered two-phase mixture (Fig.2). The lines on the squares denote
their orientations.

This structure can appear,
for example, on the intermediate scales when a random medium is formed
as a result of the stirring of the two-phase mixture.
The corresponding averaged parallel and
perpendicular conductivities of squares $\sigma_{||}(x)$ and
$\sigma_{\perp}(x)$ are defined by the following formulas
$$
\sigma_{||}(x) = x \sigma_1 + (1-x) \sigma_2 = \sigma_+ (1+ 2\epsilon z),
$$
$$
\sigma_{\perp}(x) = \left(\fr{x}{\sigma_1}+\fr{1-x}{\sigma_2}\right)^{-1}
= \sigma_+ \fr{1-z^2}{1-2\epsilon z}.
\en(35)
$$
Thus we have obtained the hierarchical representation of random medium
(in this case a two-level one). On the first level it consists  from
some regions (two different squares) of the random mixture of the two
layered conducting phases with different conductivities
$\sigma_1$ and $\sigma_2$ and arbitrary concentration. On the second level
this medium is represented as a random parquet constructed from two such
squares with different conductivities $\sigma_{||}$ and $\sigma_{\perp}$,
depending nontrivially on concentration of the initial conducting phases,
and randomly distributed with the same probabilities $p_i = 1/2$ (Fig.2).
This representation allows us to divide the averaging process into two
steps (firstly averaging over each square and then averaging over the lattice
of squares) and implement on the second step the exact formula (3).
This can be considered as some modification of the FMSA approximation.
As a result one obtains for the effective conductivity of the introduced
random two-phase model the following formula, which is applicable for
arbitrary concentration
$$
\sigma_{eff}(\epsilon,z) = \sigma_+ \sqrt{1-z^2} \tilde f(\epsilon, z),\quad
\tilde f(\epsilon, z) =
\left[\fr{1 + 2\epsilon z}{1 - 2\epsilon z}\right]^{1/2},
\en(36)
$$
This function has all necessary properties and satisfies equation (11') and
symmetry (13).
It coincides with the type (b) from (28) and has another possible functional form,
automatically satisfying the duality relation (11')
$$
\tilde f(\epsilon,z) = B(\epsilon,z)/B(-\epsilon,z)
\en(37)
$$
with a function
$B(\epsilon, z) = [1+2\epsilon z]^\fr{1}{2},$
which depends only on the combination $\epsilon z.$

It is interesting to compare this formula with the known general formulas.
In order to do this one  needs to find an asymptotic  behaviour of the derived
effective conductivity in different limiting cases.
Let us consider firstly its behaviour for small phase concentrations.

(a) In case of small concentration of the first phase $x \ll 1$ one gets
$$
\sigma_{eff}(x,z) \simeq  \sigma_2 \left(1 + \fr{2xz}{1-z^2}\right).
\en(38)
$$
It follows from (38) that an addition of small part of the first  higher
conducting phase increases an effective conductivity of the system as it
should be.

(b) In the opposite case of small concentration of the second phase
$1 - x  \ll 1$ one obtains
$$
\sigma_{eff}(x,z) \simeq  \sigma_1 \left(1 - \fr{2(1-x)z}{1-z^2}\right),
\en(39)
$$
i.e. an addition of the phase with smaller conductivity decreases
$\sigma_{eff}.$
It is worth to note that both these expressions
for arbitrary values of the conductivities $\sigma_1$ and $\sigma_2$
differ from equation (2) and coincide with it only in the weakly inhomogeneous
case $z \ll 1$.
It must be not surprising because  a form of the inclusions of the second
phase in this model has completely different, layered, structure.
In the low concentration expansion as well as
directly in formula (36) one can  see again that the divergencies appear
in the limit $z \to 1.$

(c) In case of almost equal phase concentrations
$x = 1/2 + \epsilon,$ $\epsilon \ll 1$ one obtains
$$
\tilde f(\epsilon,z) \simeq 1 + 2\epsilon z, \quad
\sigma_{eff}(\epsilon,z) \simeq \sigma_+ \sqrt{1-z^2} \left(1 +
2\epsilon z\right).
\en(40)
$$
The Keller -- Dykhne formula (3) is reproduced for equal concentrations.

The antisymmetric part of $\tilde f$ has the following form
$$
\tilde f_a(\epsilon,z) =
\fr{\tilde f^2(\epsilon,z)-1}{2\tilde f(\epsilon,z)} =
\fr{2\epsilon z}{\left[1 - 4\epsilon^2 z^2\right]^{1/2}}.
\en(41)
$$
It follows from formula (41) that for $\epsilon \ll 1$ and (or) $z \ll 1$
the odd part $f_a$ coincides in the first order with the corresponding
expression from the effective medium theory.
The corresponding function $\Phi$ is
$$
\Phi(\epsilon, z) =
\fr{\sqrt{1-z^2}}{\left[1 - 4\epsilon^2 z^2\right]^{1/2}}.
\en(42)
$$
In other words it is an analytic function of $\epsilon z$ near $\epsilon z=0$.
Basing on this formula one can conjecture that an exact expression for
$\tilde f_a(\epsilon,z)$ of this model will have a similar structure
in general case with $\sigma_i \ne 0$ or $1 - z \ne 1$.
One must note that at the same time the formula (36) does not satisfy the
equation (29) except of the trivial case $x_1 = x_2.$

For the comparison of the different expressions of the effective conductivity
we have constructed three-dimensional plots of $f(\epsilon,z)$ in
the EM approximation, in the FMSA approximation and of the
hierarchical model in FMSA-like approximation (fig.3,4).

\begin{figure}[tp]
\begin{tabular}{cc}
{\input epsf \epsfxsize=6cm \epsfbox{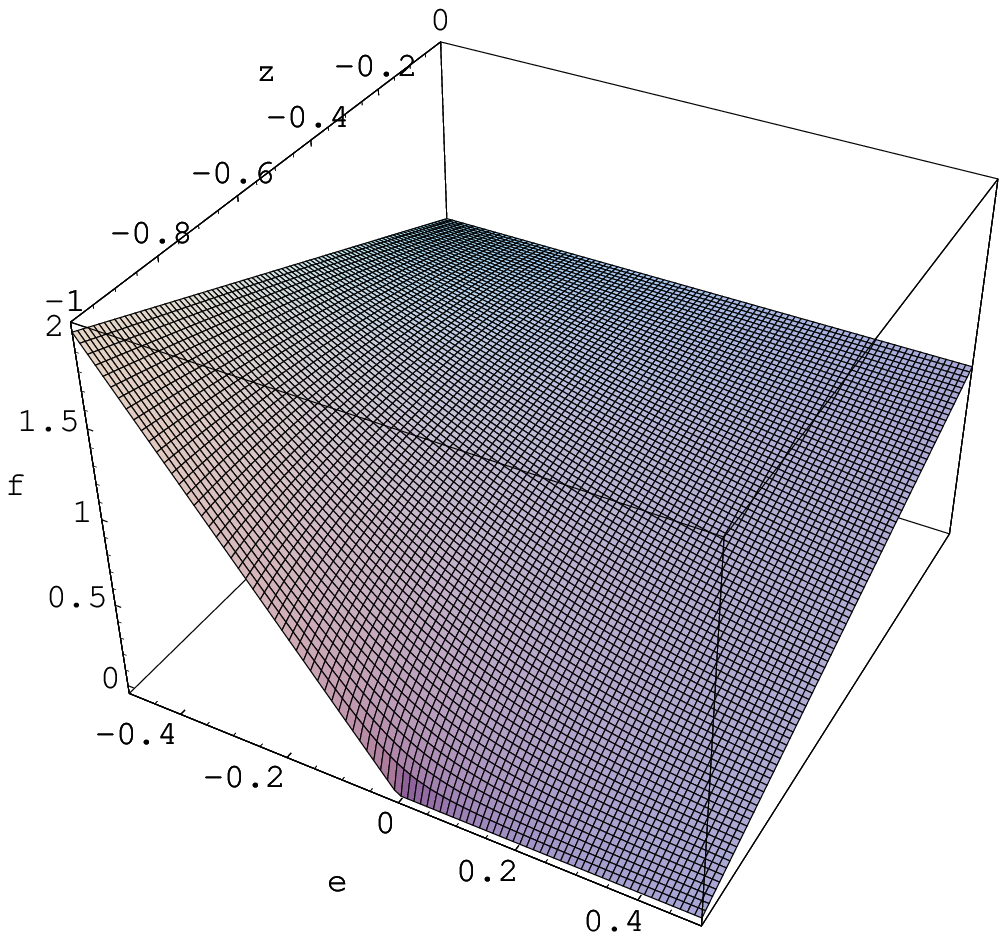}} &
{\input epsf \epsfxsize=6cm \epsfbox{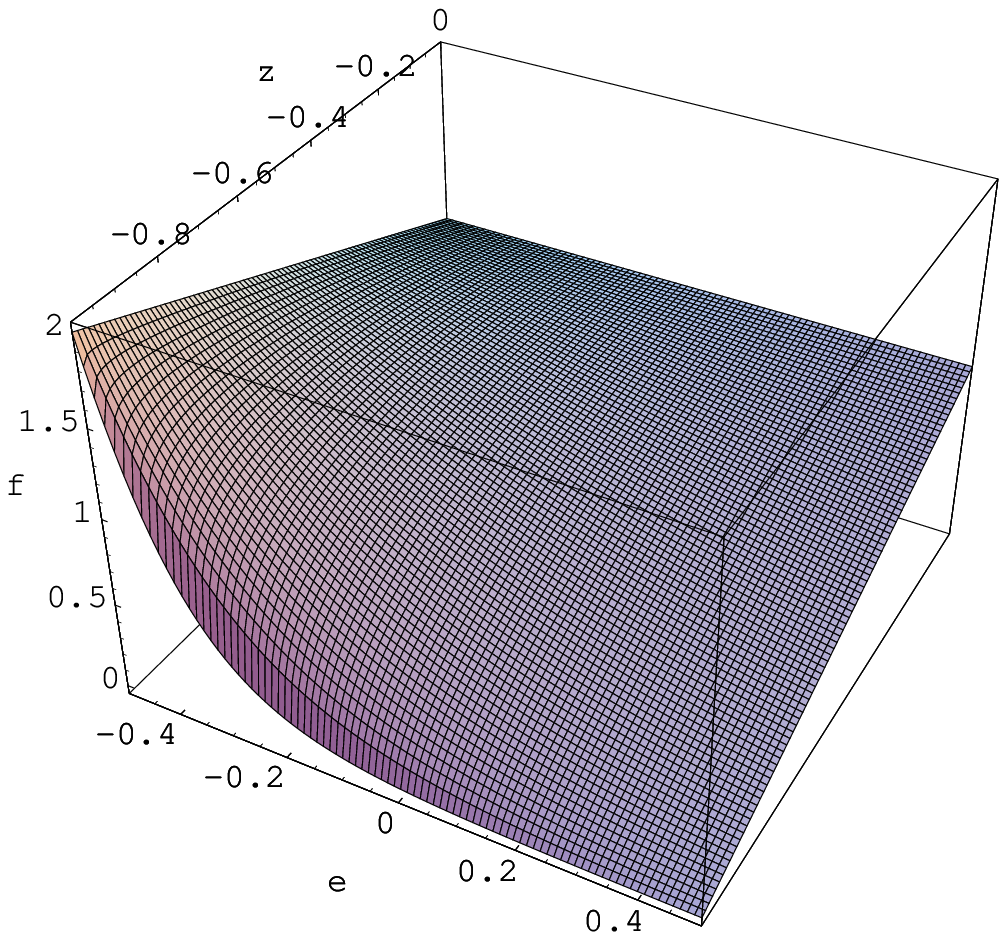}}\\
{} & {}\\
a & b\\
\end{tabular}
\vs{0.5cm}
\cl{\small Figure 3: Plots of $f(\epsilon,z)$ in :
a) EM approximation, b) FMSA approximation.}
\end{figure}
\begin{figure}
\centerline{
\input epsf \epsfxsize=7cm \epsfbox{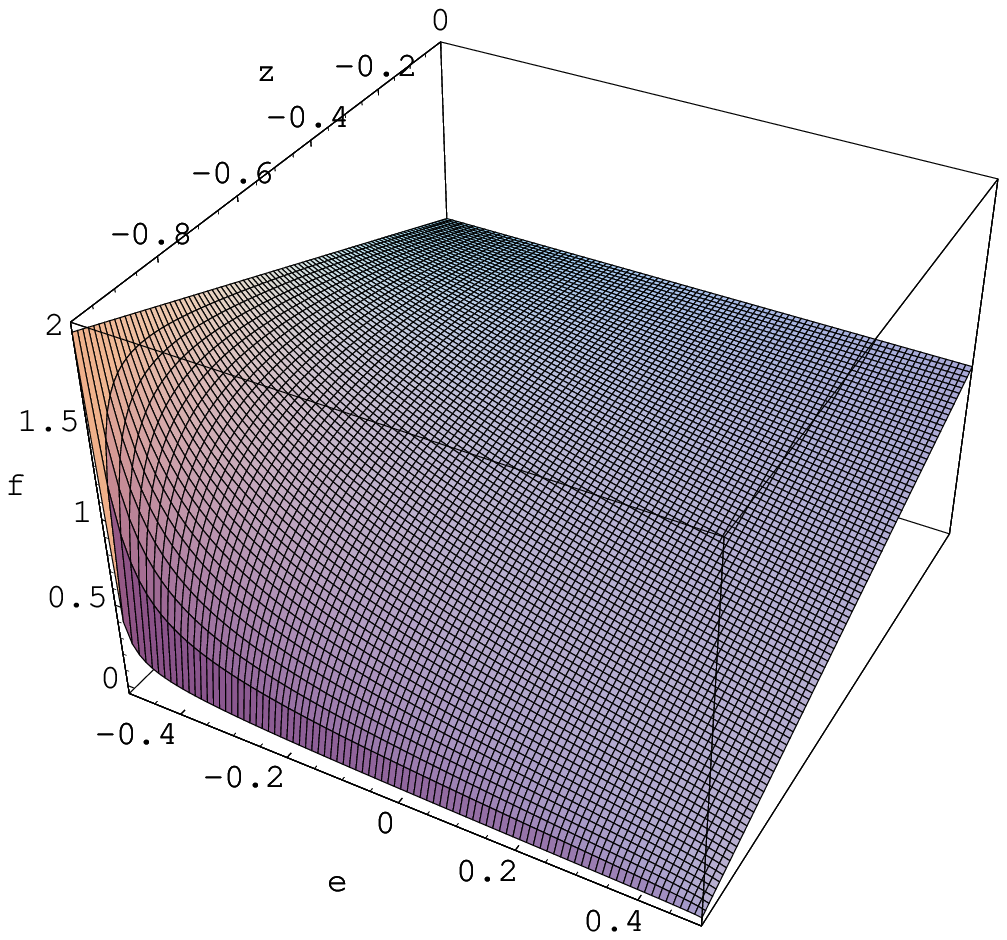}}
\vs{0.5cm}
\cl{\small Figure 4: Plot of $f(\epsilon,z)$
of composite model in FMSA-like approximation}
\end{figure}

It follows from these plots  that all three formulas for $\sigma_{eff}$,
despite of their different functional forms,
differ from each other weakly  for
$z \lesssim 0,8$ due to very restrictive boundary conditions (10') and
the exact Keller-Dykhne value. This range of $z$ corresponds approximately
to the ratio $\sigma_2/\sigma_1 \sim 10^{-1}.$ For the smaller ratios a
difference between these functions become distinguishable.

Now let us consider in the more details the derived formulas for
$\sigma_{eff}(\epsilon,z)$
in case when $\sigma_2 \to 0 (z \to 1).$ It is clear that for regularly
inhomogeneous medium one can always construct such distribution of the
conducting phase that  $\sigma_{eff}(\epsilon,1)$ will differ from zero for all
$1/2 \ge \epsilon > -1/2$.
But in the case of randomly inhomogeneous medium the limit $\sigma_2 \to 0$
is equivalent to the well known percolation problem [14,15]. In terms
of $z$ it corresponds to the limit $z \to 1$ and is also similar to the
superconducting limit $\sigma_1 \to \infty$.
Strictly speaking,  an implementation of the duality
transformation (4) is not obvious in this case.
However, if one supposes that the dual symmetry relation (4) fulfills in this
limit too due to a continuity then it follows from (4) that
$$
\sigma_{eff}(\epsilon) \sigma_{eff} (-\epsilon) = 0.
\en(43)
$$
The relation  (43) does not contradict to the known basic results
of the percolation theory that
$\sigma_{eff}(\epsilon) = 0$ for $\epsilon \le 0$ and
$\sigma_{eff}(\epsilon) \ne 0$ for $\epsilon > 0$.
Moreover it follows from the general formula (17) that in this case
$\sigma_{s} = |\sigma_a|$ and $\sigma_{eff}(\epsilon) = \sigma_a + |\sigma_a|$.
It gives
$$
\sigma_{eff}(\epsilon) =
\left\{\begin{array}{rc}
0, & \epsilon \le 0,\\
2\sigma_a, &  \epsilon > 0.\\
\end{array}\right.
\en(44)
$$
This means that a behaviour of  $\sigma_{eff}(\epsilon)$ in the
percolation theory is completely determined by its odd part.
From a general discussion of the functional structure of the conductivity
of the two-dimensional randomly inhomogeneous systems and the known
experimental and numerical results
it is known that in the percolation limit the effective conductivity
$\sigma_{eff}$ must have a
nonanalytical behaviour near the percolation
edge $x_c = 1/2$ or at small $\epsilon > 0$
$$
\sigma_{eff}(\epsilon) \sim \sigma_1 (x-x_c)^{t} \sim \sigma_1 \epsilon^t,
\en(45)
$$
where a critical exponent of the conductivity $t$ is slightly above 1
and can be represented in the form $t=1+\delta.$
Since the  values of this exponent found by the numerical calculations
are confined to be in the interval (1,10 -- 1,4) [14], then
$\delta$ have to be small and belongs to the interval (0.1 -- 0.4).
It follows from general formula (44) that one must have
$$
f_a(\epsilon, z) \xrightarrow[z \to 1, \epsilon \to 0_+]{}
\epsilon^{t}.
\en(46)
$$
It means that the function $\Phi(\epsilon,z)$ has at small $\epsilon$
some crossover on $z$ under $z \to 1$ from a regular (analytical)
behaviour to a singular one. At the moment an exact form of this crossover
is unknown. For example, it can be of the form
$$
\Phi(\epsilon,z) \sim
\left(\Phi_0(z^2) + \Phi_1(z^2) \epsilon^2\right)^{\delta(z)/2},
\quad z \to 1,
\en(47)
$$
where $\Phi_0(z^2) \to 0,$ $\Phi_1(z^2) \to \Phi_1 \ne 0$ and
$\delta(z) \to \delta \ne 0$ when $z \to 1.$

However, as it follows from the formulas obtained above,
one gets always  $\sigma_{eff} \to 0$ in the limit  $\sigma_2 \to 0$, except
the region near $x=1.$
It means that all these formulas obtained in FMSA approximation
are not valid in the limit
$\sigma_2 \to 0.$ This is confirmed by the appearance of the divergencies
in the expansions of $\sigma_{eff}$ in small concentrations in the limit
$z \to 1.$
This fact is a consequence of the made approximation.
For example, in case of the model of the layered squares this is due to the
"closing" or "locking" effect of the layered structure
in the adopted approximation in the limit $\sigma_2 \to 0.$ In order
to obtain a finite conductivity in this model above threshold concentration
$x_c$ one needs to take into account the correlations between adjacent squares.
It is easy to show that near the threshold an effective conductivity is
determined by random conducting clusters formed out of the crossing
random layers from neighboring elementary squares. As is well known,
the mean size of these clusters
diverges near the percolation threshold \ci{14,15}
and for this reason the FMSA approximation cannot be
applicable for the description of $\sigma_{eff}$ in the limit
$z\to 1 \, (\epsilon > 0)$ and of
the percolation problem. It follows from figs. 3 that the EM approximation
overestimates $\sigma_{eff}$, and both other formulas underestimate it
in the region $z \to 1, \epsilon > 0.$ We hope to investigate this limit in
detail in the subsequent papers.

Thus we have discussed the possible functional forms of the effective
conductivity of the two-phase system at arbitrary values of concentrations.
A new functional equation, generalizing the duality relation,
was deduced in the FMSA approximation and
its solution was found. We have constructed  also a hierarchical model of
the random inhomogeneous
medium and have found its effective conductivity in the same approximation
at arbitrary phase concentrations. All  formulas for the effective
conductivity have the different functional forms. They

(1) satisfy all symmetries including the dual symmetry and all necessary
inequalities,

(2) reproduce the general formulas for $\sigma_{eff}$ in the weakly
inhomogeneous case.

All these results confirm a conjecture that, in general, $\sigma_{eff}$ of the
two-phase randomly inhomogeneous system may be a nonuniversal function and
can depend on some details of the structure of the randomly inhomogeneous
regions.
Analogous conclusions were done in paper \ci{16} during a discussion
of the possibility to find the generalization of the formula (3) for case
$N \ge 3$.

The obtained formulas can be used for the approximation of the effective conductivity
of some real randomly inhomogeneous systems like a corresponding formula
of the EM approximation.
The introduced composite method of the construction of the model
random medium
can be generalized on the other ways of determination of the effective
intermediate conducting boxes. It can be done for different types of boxes
as well as for different numbers of the possible types of the boxes.
It is clear that then one will have to use another formulas instead of (3).

This work was supported be RFBR grants 00-15-96579 and 02-02-16403.

\end{document}